\documentclass[english]{aastex62}
\usepackage{array}
\usepackage{booktabs}
\usepackage{mathtools}
\usepackage{amsbsy}
\usepackage{amstext}
\usepackage{amssymb}
\usepackage{cancel}
\usepackage{mathdots}
\usepackage{stmaryrd}
\usepackage{stackrel}
\usepackage{graphicx}
\PassOptionsToPackage{version=3}{mhchem}
\usepackage{mhchem}
\usepackage{esint}

\makeatletter

\providecommand{\tabularnewline}{\\}

\pdfoutput=1
\usepackage{babel}

\makeatother

\usepackage{babel}
\begin{document}

\title{DOES NONAXISYMMETRIC DYNAMO OPERATE IN THE SUN?}

\author{V.V. Pipin}

\affiliation{Institute of Solar-Terrestrial Physics, Irkutsk, 664033,
Russian Federation}

\author{A.G. Kosovichev}

\affiliation{Center for Computational Heliophysics, New Jersey Institute
of Technology, Newark, NJ 07102, USA} \affiliation{Department of
Physics, New Jersey Institute of Technology, Newark, NJ 07102, USA} 
\begin{abstract}
We explore effects of random non-axisymmetric perturbations of kinetic
helicity (the $\alpha$ effect) and diffusive decay of bipolar magnetic
regions on generation and evolution of large-scale non-axisymmetric
magnetic fields on the Sun. Using a reduced 2D nonlinear mean-field
dynamo model and assuming that bipolar regions emerge due to magnetic
buoyancy in situ of the large-scale dynamo action, we show that fluctuations
of the $\alpha$ effect can maintain the non-axisymmetric magnetic
fields through a solar-type $\alpha^{2}\Omega$ dynamo process. It
is found that diffusive decay of bipolar active regions is likely
to be the primary source of non-axisymmetric magnetic fields observed
on the Sun. Our results show that non-axisymmetric dynamo models
with stochastic perturbations of the $\alpha$ effect can explain
periods of extremely high activity (`super-cycle' events) as well
as periods of deep decline of magnetic activity. We compare the models
with synoptic observations of solar magnetic fields for the last four
activity cycles, and discuss implications of our results for interpretation
of observations of stellar magnetic activity. 
\end{abstract}


\section{Introduction}

Since the seminal papers of \citet{choud92} and \citet{h93}, random
variations of kinetic helicity in dynamo processes (the so-called
$\alpha$ effect) are often considered as the main source of long-term
variations of solar activity cycles \citep{oss-h96a,moss-sok08,uetal09,pipea2012AA,2014AA563A18P}.
In the standard mean-field framework, turbulent generation of magnetic
fields results from reflection-symmetry breaking of helical convection
motions \citep{KR80}. In the mean-field theory, the effect is described
by the mean-electromotive force, 
\[
\mathbf{\mathcal{E}}=\left\langle \mathbf{u}\times\mathbf{b}\right\rangle ={\alpha}\circ\left\langle \mathbf{B}\right\rangle +\dots,
\]
where $\mathbf{u}$ is the turbulent velocity, $\mathbf{b}$ is the
turbulent magnetic field, $\left\langle \mathbf{B}\right\rangle $
is the large-scale magnetic field, and coefficient $\alpha=-\frac{1}{3}\left\langle\mathbf{u}\cdot\nabla\times\mathbf{u}\right\rangle\tau_{{\rm cor}}$
is a pseudo-scalar proportional to the kinetic helicity, $\mathbf{u}\cdot\nabla\times\mathbf{u}$,
and turbulent correlation times, $\tau_{{\rm cor}}$. The amount of
convective energy, which can be spent on turbulent generation of
the large-scale magnetic field by the $\alpha$ effect, is only few
percents of the total convective energy \citep{park}. Taking this
constraint into account, it was shown that magnitude of the $\alpha$
effect can randomly vary in each hemisphere in the range from 10 to
20 percent (see, \citealt{choud92,h93,moss-sok08}). However, the
results of \citet{choud92} and \citet{oss-h96a} showed that in order
to explain the Grand cycles of solar magnetic activity the random
fluctuations should be of the same order as the mean magnitude of
the $\alpha$-effect. Results of \citet{moss-sok08} showed that the
time scale of fluctuations should also be taken into account. They
found that if the correlation time is comparable to the cycle duration
then the fluctuations with amplitude of few dozen percents are sufficient
to explain the Grand minima of solar activity.

The current paradigm assumes that sunspots are formed from large-scale
axisymmetric toroidal magnetic field emerging from the solar convection
zone where the field is regenerated by hydromagnetic dynamo. Results of \citet{KR80} and \citet{rad86AN}
showed that, because of the differential rotation, the solar dynamo
can not maintain a regular non-axisymmetric large-scale magnetic field.
Nevertheless, large-scale non-axisymmetric magnetic fields are commonly
observed on the Sun, for example, in the form of coronal holes \citep{1974Natur250.717G},
which represent regions of open magnetic flux \citep{stix77}. The
surface flux-transport models successfully simulate the process of
formation of coronal holes from decaying active regions \citep{1990ApJ355.726W,2017ApJ843.111C}.
These models assume that influence of surface non-axisymmetric magnetic
fields on the dynamo action in the deep convection zone is negligible.
{\citet{moss99} and \citet{bigruz} showed that weak large-scale non-axisymmetric field structures may be consistent with non-linear
mean field models of the solar dynamo, in which non-axisymmetric dynamo modes are maintained by either non-linearity of $\alpha$-effect quenching, or non-axisymmetric distribution of $\alpha$. It was suggested that the excitation of the non-axisymmetric modes can be sensitive to the radial dependence of the rotation law \citep{moss99,2017MNRAS.466.3007P}.  \citet{pk15} studied response of a nonlinear non-axisymmetric mean-field solar dynamo model to non-axisymmetric perturbations and showed that the effect can depend on the root depth of the non-axisymmetric magnetic fields. The  non-axisymmetric dynamo models may be relevant to
the problem of solar active longitudes \citep{berd06}.} Observational results of \citet{2012AA547A93S} showed that the presence of background (basal) magnetic flux observed on the surface of the Sun does not depend on the magnetic cycle. From this consideration it seems that much of the basal flux may well originate from the global dynamo. This flux may persists during the solar minima because diffusion of solar bipolar regions could take long time.

In this paper we explore additional possibility which stems from non-axisymmetric
dynamo action. Longitudinal fluctuations of the $\alpha$ effect are
usually ignored in mean-field stellar dynamo models. Using nonlinear
mean-field dynamo models we show that such non-axisymmetric random
perturbations can maintain large-scale non-axisymmetric magnetic fields
in a solar-type $\alpha^{2}\Omega$ dynamo. Our goal is to investigate
the process of stochastic excitation of large-scale non-axisymmetric
magnetic field and estimate how it affects the large-scale basal magnetic
flux. To compare this mechanism with diffusive decay of bipolar active
regions, we simulate active region emergence using the Parker's magnetic
buoyancy effect. To demonstrate the difference between the two competitive
mechanisms we employ a reduced 2D non-linear and non-axisymmetric
dynamo model which describes the dynamo wave propagation on the spherical
surface. The modeling results are compared with synoptic observations
of large-scale solar magnetic fields on the Sun.

The paper is organized as follows. Section 2 describes the dynamo
model and its parameters. Section 3 presents results of numerical
simulations for various model conditions. Section 4 gives an outline
of observational data and comparison with the model. The final section
summarizes and discusses the main results of our analysis.

\section{Nonaxisymmetric 2D Dynamo Model}

\subsection{Governing Equations}

In the framework of mean-field magnetohydrodynamics \citep{KR80}
the evolution of the large-scale magnetic field, $\left\langle \mathbf{B}\right\rangle $,
in perfectly conductive media is governed by the induction equation,
\begin{equation}
\partial_{t}\left\langle \mathbf{B}\right\rangle =\boldsymbol{\nabla}\times\left(\boldsymbol{\mathcal{E}}+\left\langle \mathbf{U}\right\rangle \times\left\langle \mathbf{B}\right\rangle \right),\label{eq:mfe-1}
\end{equation}
where, $\boldsymbol{\mathcal{E}}=\left\langle \mathbf{u\times b}\right\rangle $
is the mean electromotive force with $\mathbf{u}$ and $\mathbf{b}$
standing for the turbulent velocity and magnetic field respectively.
It is convenient \citep{1990AA230.463J} to represent the large-scale
magnetic field induction vector in terms of the axisymmetric and non-axisymmetric
components as follows: 
\begin{eqnarray}
\left\langle \mathbf{B}\right\rangle  & = & \overline{\mathbf{B}}+\tilde{\mathbf{B}}\label{eq:b0}\\
\mathbf{\overline{B}} & = & \hat{\boldsymbol{\phi}}B+\nabla\times\left(A\hat{\boldsymbol{\phi}}\right)\label{eq:b1}\\
\tilde{\mathbf{B}} & = & \boldsymbol{\nabla}\times\left(\mathbf{r}T\right)+\boldsymbol{\nabla}\times\boldsymbol{\nabla}\times\left(\mathbf{r}S\right),\label{eq:b2}
\end{eqnarray}
where $\overline{\mathbf{B}}$ and $\tilde{\mathbf{B}}$ are the axisymmetric
and non-axisymmetric components; ${A}$, ${B}$, ${T}$ and ${S}$
are scalar functions representing the field components; $\hat{\boldsymbol{\phi}}$
is the azimuthal unit vector, $\mathbf{r}$ is the radius vector;
$r$ is the radial distance, and $\theta$ is the polar angle. Hereafter,
the overbar denotes the axisymmetric magnetic field, and tilde denotes
non-axisymmetric properties.

To elucidate basic properties of the non-axisymmetric dynamo action,
we consider a reduced dynamo model in which the radial dependence
of the magnetic field is disregarded. In this case, the induction
vector of the large-scale magnetic field is represented in terms of
the scalar functions as follows: 
\begin{eqnarray*}
\left\langle \mathbf{B}\right\rangle  & = & -\frac{\mathbf{r}}{R^{2}}\frac{\partial\sin\theta A}{\partial\mu}-\frac{\hat{\theta}}{R}A+\hat{\boldsymbol{\phi}}B\\
 & - & \frac{\mathbf{r}}{R^{2}}\Delta_{\Omega}S+\frac{\hat{\theta}}{\sin\theta}\frac{\partial T}{\partial\phi}+\hat{\phi}\sin\theta\frac{\partial T}{\partial\mu},
\end{eqnarray*}
where $R$ represents the radius of the spherical surface inside a
star where the hydromagnetic dynamo operates. The model employs the
following expression of $\boldsymbol{\boldsymbol{\mathcal{E}}}$:

\begin{eqnarray}
\boldsymbol{\boldsymbol{\mathcal{E}}} & = & \alpha\circ\left\langle \mathbf{B}\right\rangle -\eta_{T}\boldsymbol{\nabla}\times\mathbf{\left\langle B\right\rangle }+V_{\beta}\hat{\mathbf{r}}\times\mathbf{B}.\label{eq:simpE}
\end{eqnarray}
Here, it is assumed that: 
\begin{equation}
\text{\ensuremath{\alpha}}_{ij}=\alpha_{0}\psi\left(\left|\left\langle \mathbf{B}\right\rangle \right|\right)\cos\theta\delta_{ij},\label{eq:alft}
\end{equation}
where coefficient $\alpha_{0}$ represents the magnitude of the $\alpha$-effect,
and $\psi$ is the standard magnetic quenching function \citep{pi08Gafd}.
The magnitude, $\alpha_{0}$, is allowed to vary randomly in time
$t$ and longitude $\phi$:

\begin{equation}
\alpha_{0}=\overline{\alpha}\left(1+\xi_{\alpha}\left(\phi,t\right)\right),\label{eq:alf}
\end{equation}
where $\overline{\alpha}$ is the stationary axisymmetric part of
$\alpha$-effect; the function $\xi_{\alpha}\left(\phi,t\right)$
describes random fluctuations. Details of the fluctuations will be
defined further. The second term represents turbulent diffusion with
coefficient $\eta_{T}$. The last term in Eq (\ref{eq:simpE}) describes
the magnetic buoyancy effect. Here, $\hat{\mathbf{r}}$ is the radial
unit vector, and $V_{\beta}$ is the escape velocity of magnetic field,
The escape velocity, $V_{\beta}$, accounts for the loss of generated
magnetic flux from the dynamo region \citep{1984ApJ...281..839P,1984ApJ...287..769N,1990AA240.142M}.
\citet{park} suggested that the magnetic buoyancy can result in formation
of bipolar active regions. Currently, it becomes evident that the
magnetic buoyancy is not the only mechanism forming emerging active
regions of the Sun \citep{2001ARep...45..569G,2010ApJ719.307K,2012ApJ753L13S,2013ApJ762.130L,2013ApJ776L23B,2017IAUS327.46L,2018FrASS517M}.
In this study, we assume that the magnetic buoyancy acts on relatively
small-scale parts of the axisymmetric magnetic field, perhaps, because
of some kind of nonlinear instability, and contributes to generation
of the non-axisymmetric magnetic field component. It is formulated
following \citet{kp93}: 
\begin{eqnarray}
V_{\beta}=\begin{cases}
\dfrac{\alpha_{MLT}u'}{\gamma}\beta^{2}K\left(\beta\right)\left[1+\xi_{\beta}\left(\phi\right)\right], & \text{if}\;\beta\ge\beta_{cr},\\
0, & \text{if}\;\beta<\beta_{cr}
\end{cases}\label{eq:buoy}
\end{eqnarray}
where $\beta=\left|\left\langle \mathbf{B}\right\rangle \right|/\mathrm{B_{eq}}$,
$\mathrm{B_{eq}}=\sqrt{4\pi\overline{\rho}u'^{2}}$, function $K\left(\beta\right)$
is defined in \citet{kp93}, function $\xi_{\beta}\left(\phi\right)$
describes the longitudinal dependence of the instability, and parameter
$\beta_{cr}$ controls the instability threshold. These parameters
will be described below. From results of the above cited paper, it
follows that for $\beta\ll1$, $K\left(\beta\right)\sim1$, and for
$\beta>1$, $K\left(\beta\right)\sim1/\beta^{3}$. In this formulation,
the preferable latitude of the ``active region emergence'' is determined
by maximum of the toroidal magnetic field energy, see Eq.(\ref{eq:buoy}).
Parameter $\beta_{cr}=0.5$ is used to prevents emergence of active
regions at high latitudes.

The minimal set of the dynamo equations to model the non-axisymmetric
magnetic field evolution can be obtained by generalization of the
1D model suggested by \citet{park93}, which has a solution in the
form of dynamo waves migrating towards the equator. The model studied
extensively, for example, by \citet{ku98,moss-sok08,uetal09}. In
this framework, the radial dependence of magnetic field is disregarded,
and it is assumed that the radial gradient of angular velocity is
greater than the latitudinal gradient. Applying these simplifications
to Eq (\ref{eq:mfe-1}) and Eqs (\ref{eq:b0}-\ref{eq:b2}) we obtain
the following set of dynamo equations in terms of the scalar functions,
$A,B,S$, and $T$: 
\begin{eqnarray}
\partial_{t}B & =- & \sin\theta\frac{\partial\Omega}{\partial r}\frac{\partial\left(\sin\theta A\right)}{\partial\mu}+\eta_{T}\frac{\sin^{2}\theta}{R^{2}}\frac{\partial^{2}\left(\sin\theta B\right)}{\partial\mu^{2}}-\frac{B}{\tau}\label{eq:bt}\\
 &  & +\frac{\sin\theta}{R}\frac{\partial}{\partial\mu}\alpha_{0}\mu\left\langle B_{r}\right\rangle +\frac{\alpha_{0}\mu}{R}\left\langle B_{\theta}\right\rangle -\frac{1}{R}V_{\beta}\left\langle B_{\phi}\right\rangle \nonumber 
\end{eqnarray}
\begin{equation}
\partial_{t}A=\alpha_{0}\mu\left\langle B_{\phi}\right\rangle +\eta_{T}\frac{\sin^{2}\theta}{R^{2}}\frac{\partial^{2}\left(\sin\theta A\right)}{\partial\mu^{2}}-\frac{V_{\beta}}{R}A-\frac{A}{\tau},\label{eq:at}
\end{equation}
\begin{eqnarray}
\partial_{t}\Delta_{\Omega}T+\Delta_{\Omega}\delta\Omega\frac{\partial T}{\partial\phi} & = & -\frac{1}{R}\frac{\partial\Omega}{\partial r}\sin^{2}\theta\frac{\partial\Delta_{\Omega}S}{\partial\mu}+\frac{\eta_{T}}{R^{2}}\Delta_{\Omega}^{2}T\label{eq:Tt}\\
 &  & +\Delta_{\Omega}\frac{\alpha_{0}\mu}{R}\left(\left\langle B_{r}\right\rangle \sin^{2}\theta+\mu\sin\theta\left\langle B_{\theta}\right\rangle \right)-\frac{1}{R}\frac{\partial}{\partial\phi}\left[\frac{\alpha_{0}}{\sin\theta}\mu\left\langle B_{\phi}\right\rangle \right]\nonumber \\
 & + & \frac{1}{R}\frac{\partial}{\partial\mu}\alpha_{0}\mu\sin\theta\left\{ \mu\sin\theta\left\langle B_{r}\right\rangle +\mu^{2}\left\langle B_{\theta}\right\rangle \right\} \nonumber \\
 & - & \frac{1}{R\sin\theta}\frac{\partial}{\partial\phi}\left\langle B_{\theta}\right\rangle V_{\beta}-\frac{\partial}{\partial\mu}\left(\sin\theta\left\langle B_{\phi}\right\rangle V_{\beta}\right),\nonumber 
\end{eqnarray}

\begin{eqnarray}
\partial_{t}\Delta_{\Omega}S+\left(\delta\Omega\Delta_{\Omega}\frac{\partial}{\partial\phi}S\right) & = & \frac{\eta_{T}}{R^{2}}\Delta_{\Omega}^{2}S+\frac{\partial}{\partial\mu}\alpha_{0}\mu\sin\theta\left\langle B_{\phi}\right\rangle \label{eq:St}\\
 & + & \frac{\partial}{\partial\phi}\left\{ \frac{\alpha_{0}\mu}{\sin\theta}\left(\left\langle B_{\theta}\right\rangle +\sin\theta\left(\mathbf{e}\cdot\left\langle \mathbf{B}\right\rangle \right)\right)\right\} \nonumber \\
 & - & \frac{1}{\sin\theta}\frac{\partial}{\partial\phi}\left(\left\langle B_{\phi}\right\rangle V_{\beta}\right)+\frac{\partial}{\partial\mu}\left(\sin\theta\left\langle B_{\theta}\right\rangle V_{\beta}\right),\nonumber 
\end{eqnarray}
where ${\displaystyle \Delta_{\Omega}=\frac{\partial}{\partial\mu}\sin^{2}\theta\frac{\partial}{\partial\mu}+\frac{1}{\sin^{2}\theta}\frac{\partial^{2}}{\partial\phi^{2}}}$
and $\mu=\cos\theta$. The $\tau$-terms in Eqs(\ref{eq:bt},\ref{eq:at})
were suggested by \citet{moss-sok08} to account for turbulent diffusion
in radial direction. {Similarly to the cited paper we put ${\displaystyle \tau=3\frac{R^{2}}{\eta_{T}}}$.
}The $\tau$-parameter will be specified in the next subsection. For
brevity, some terms in above equations are given explicitly via components
of the magnetic induction vector, $\left\langle \mathbf{B}\right\rangle $.
These terms result in coupling between different modes of the large-scale
magnetic field in the case of longitudinal dependence of parameters
$\alpha_{0}$ and $V_{\beta}$. If the differential rotation is strong
then the non-axisymmetric modes are stable against the dynamo instability.
The nonlinear coupling (e.g., due to the $\alpha$-effect) is not
very effective for maintaining non-axisymmetric dynamo modes \citep{bran89,radler90,moss99,pk15}.
To simulate stretching of non-axisymmetric magnetic field by the surface
differential rotation we consider the latitudinal dependence of angular
velocity $\delta\Omega=-0.25\sin^{2}\theta\Omega$ in Eqs (\ref{eq:Tt})
and (\ref{eq:St}), which are written in the coordinate system rotating
with angular velocity $\Omega$. {In this
paper, we discuss relatively simple 2D models with the constant rotational shear. }

\begin{figure}
\centering \includegraphics[width=0.45\columnwidth]{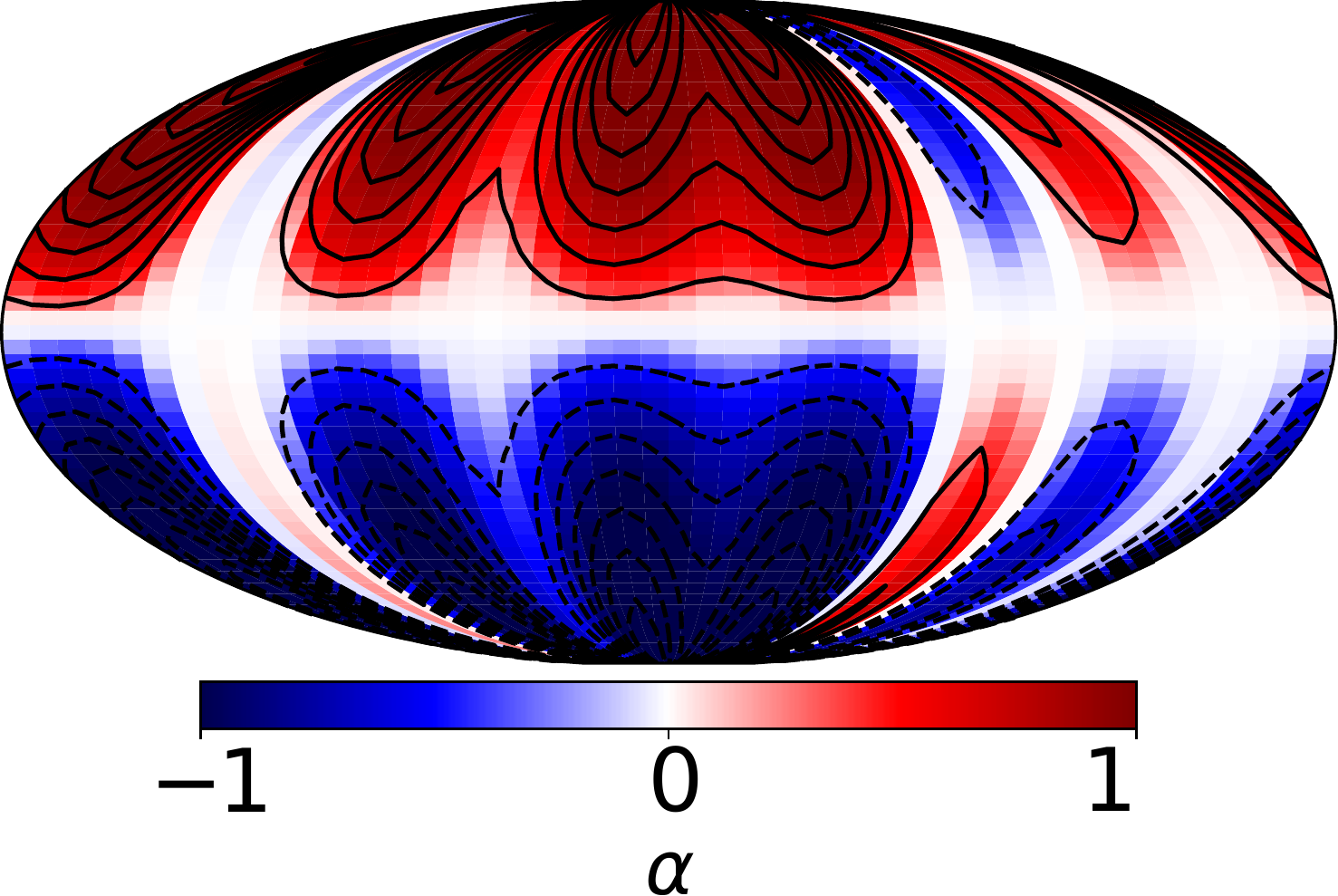} \caption{\label{fig:alpran}Snapshot of a random realization of the $\alpha$
effect with $\sigma_{\xi}=0.25$ (see, Sec. \ref{subsec:Stochastic-parameters}).
Contours are drawn in range of $\pm2$.}
\end{figure}

The numerical scheme employs a pseudo-spectral approach for integration
along latitude. For the non-axisymmetric components, we employ the
spherical harmonic decomposition, i.e., scalar functions $T$ and
$S$ are represented in the form: 
\begin{eqnarray}
T\left(\mu,\phi,t\right) & = & \sum\hat{T}_{\ell,m}\left(t\right)\bar{P}_{\ell}^{\left|m\right|}\exp\left(-im\phi\right),\label{eq:tdec}\\
S\left(\mu,\phi,t\right) & = & \sum\hat{S}_{\ell,m}\left(t\right)\bar{P}_{\ell}^{\left|m\right|}\exp\left(-im\phi\right),\label{eq:sdec}
\end{eqnarray}
where $\bar{P}_{\ell}^{m}$ is the normalized associated Legendre
function of degree $\ell\ge1$ and azimuthal order $m\ge1$. Note
that $\hat{S}_{l,-m}=\hat{S}_{l,m}^{*}$, and the same is valid for
$\hat{T}$. Our typical simulation runs were performed for 990 spherical
harmonics (with $\ell_{max}=36$, $m_{max}=18$). In addition, some
runs were performed with higher resolution including 1752 spherical
harmonics ($\ell_{max}=48$, $m_{max}=24$). All the nonlinear terms
are calculated explicitly in the real space. The numerical integration
is carried out in latitude from pole to pole.

To quantify asymmetry of the mean radial magnetic field distribution
relative to the equator, we introduce the parity index, $P_{E}$:
\begin{eqnarray}
P_{E} & = & \frac{E_{q}-E_{d}}{E_{q}+E_{d}},\label{eq:par}\\
E_{d} & = & \frac{1}{4}\int\left(\overline{B}_{r}\left(\mu\right)-\overline{B}_{r}\left(-\mu\right)\right)^{2}d\mu,\nonumber \\
E_{q} & = & \frac{1}{4}\int\left(\overline{B}_{r}\left(\mu\right)+\overline{B}_{r}\left(-\mu\right)\right)^{2}d\mu,\nonumber 
\end{eqnarray}
where $E_{d}$ and $E_{q}$ are the energy of the antisymmetric and
symmetric modes, respectively.

We consider decomposition of the radial magnetic field into dynamo
modes: 
\[
\left\langle \mathbf{B}_{r}\right\rangle =\sum B_{r}^{(m)}\left(\mu\right)\mathrm{e^{im\phi}},
\]
where the case of $m=0$ corresponds to the axisymmetric magnetic
field. The degree of non-axisymmetry of the magnetic field is described
by the ratio: 
\begin{eqnarray}
P_{X} & = & {\displaystyle \frac{\tilde{E}_{r}^{(m)}}{E_{r}^{(m)}},}\label{eq:nx}\\
\tilde{E}_{r}^{(m)} & = & \frac{1}{8\pi}\sum_{m\ge1}\intop B_{r}^{(m)}B_{r}^{(m)*}d\mu,\label{eq:rm}
\end{eqnarray}
and $E_{r}^{(m)}$ is the total energy of the radial component of
magnetic field. Other characterization parameters are the mean strength
of the unsigned radial magnetic field: $\overline{\left|\left\langle \mathbf{B}_{r}\right\rangle \right|}=\sqrt{E_{r}^{(m)}/2\pi}$,
and the unsigned toroidal magnetic field: $\overline{\left|\left\langle \mathbf{B}_{\phi}\right\rangle \right|}=\sqrt{E_{\phi}^{(m)}/2\pi}$,
where $E_{\phi}^{(m)}$ is the total energy of the toroidal component.
The similar parameters are introduced for the axisymmetric magnetic
field.

\subsection{Nonaxisymmetric perturbations\label{subsec:Stochastic-parameters} }

Variations of the $\alpha$-effect are modeled by Eq.~(\ref{eq:alf}).
The time scales in our model is defined in terms the characteristic
diffusion time, ${\displaystyle \frac{R^{2}}{\eta_{T}}}$. In these
units, the basic dynamo period (corresponding to the 11-year cycle)
is $P\approx0.15{\displaystyle \frac{R^{2}}{\eta_{T}}}$. The fluctuations
are imposed randomly in time. We consider both, short and long correlation
times. For the solar case, the short correlation time, $\tau_{\xi}=0.02P$,
is roughly equal to the correlation time of non-axisymmetric modes
of solar magnetic fields \citep{pial14}. 
In addition, we consider the long correlation time, $\tau_{\xi}=0.5P$.
The longitudinal dependence of function $\xi_{\alpha}\left(\phi,t\right)$
is modeled as a superposition of of sinusoidal oscillations with random
amplitudes and phases. The strength of fluctuations is controlled
by parameter $\sigma_{\xi}$ which is the standard deviation (STD)
of the random variable, $\xi_{\alpha}$. As in the model of \citet{ps11},
we consider a set of $\sigma_{\xi}=0.25,0.5,1$ for the short and
long correlation time $\tau_{\xi}$ cases. The longitudinal fluctuations
of the $\alpha$ effect are modeled as a superposition of random harmonics
up to the octupole modes.

In practice, the fluctuations of the $\alpha$ effect are implemented
as follows. We use the pseudo-random number generator library of SciPy
(scipy.org) to produce a Gaussian set of random intervals with correlation
time $\tau_{\xi}$. Then, during the dynamo run, the distribution
of $\xi_{\alpha}\left(\phi,t\right)$ is calculated for each of the
random time intervals. In this step, we generate a random Gaussian
sequence of length $N_{\phi}$ (the number of mesh points in azimuth)
with the STD of $N_{\phi}\sigma_{\xi}$. Then, using the fast Fourier
transform we filter out all harmonics higher than the octupole. The
obtained distribution of $\xi_{\alpha}\left(\phi,t\right)$ is used
in the model until the next random fluctuation. A snapshot of a random
realization of $\alpha$ is illustrated in Figure \ref{fig:alpran}.

Magnetic buoyancy instability perturbations are determined by function:
\begin{equation}
\xi_{\beta}\left(\phi\right)=C_{\beta}\exp\left(-m_{\beta}\sin^{2}\left(\frac{\phi-\phi_{0}}{2}\right)\right).\label{eq:xib}
\end{equation}
The instability is randomly initiated in the northern or southern
hemispheres, and the longitude, $\phi_{0}$, is also chosen randomly.
We arbitrary chose the fluctuation interval $\tau_{\beta}=0.01P$.
After injection of the perturbation the evolution is solely determined
by the dynamo equations. Parameter $m_{\beta}$ controls the spatial
scale of the instability. Theoretically, using a high values of $m$
we can reproduce the spatial scale of the solar active regions. However,
this requires increasing the resolution in both longitude and latitude,
and becomes computationally expensive. We chose the value $m_{\beta}=50$,
which allows us to accurately resolve the evolving non-axisymmetric
perturbations of magnetic field, and qualitatively reproduce the essential
physical effects. Parameter $C_{\beta}$ controls the amount of the
injected magnetic flux. If large-scale toroidal magnetic flux at a
given co-latitude $\theta$ is transformed into magnetic flux of the
perturbation then $\left\langle \xi_{\beta}\left(\phi,t\right)\right\rangle _{\phi}\approx1$.
This condition corresponds to $C_{\beta}\approx15$. In reality, solar
active regions are formed by concentration of the toroidal magnetic
flux emerging in the photosphere. Turbulent convective motions and
other physical processes may take part in the process of formation
of solar active regions. Therefore, parameter $C_{\beta}$ can be
higher than the above mentioned value, and we choose it about three
times greater $C_{\beta}=40$ than the value corresponding to the
local large-scale field. In the numerical experiments, we found that
higher values of $C_{\beta}$ result in strong cycle-to-cycle variability
of the magnetic energy even in the case of stationary $\alpha$-effect.
For the chosen $C_{\beta}$, $\left|\left\langle \mathbf{B}\right\rangle \right|\approx2\left|\overline{B_{\phi}}\right|$,
therefore fluctuations of the magnetic field because of the magnetic
buoyancy instability are of the order of the axisymmetric toroidal
magnetic field strength, which is widely accepted in the literature
\citep{KR80,brsu05}. This part of the model can be improved using
the bipolar region production algorithms from the flux-transport models.

\begin{table}
\centering \caption{\label{tab:C}Parameters of the dynamo models, $P$ is the dynamo
period.}
\begin{tabular}{>{\centering}p{1cm}>{\centering}p{1cm}>{\centering}p{2cm}>{\centering}p{1.5cm}}
\toprule 
Model  & $C_{\beta}$  & $\sigma_{\xi}$  & $\tau_{\xi}/P$\tabularnewline
\midrule 
M0  & $0$  & 0  & -\tabularnewline
\midrule 
M1a  & 40  & 0  & -\tabularnewline
\midrule 
M1b  & 0  & 0.25  & 0.02\tabularnewline
\midrule 
M2a  & 40  & 0.25  & $0.02$\tabularnewline
\midrule 
M2b  & 40  & 0.5  & $0.02$\tabularnewline
\midrule 
M2c  & 40  & 1.  & $0.02$\tabularnewline
\midrule 
M2d  & 40  & 1. ,

$\tilde{\xi}_{\alpha}=0$  & $0.02$\tabularnewline
\midrule 
M3a  & 40  & 0.25  & $0.5$\tabularnewline
\midrule 
M3b  & 40  & 0.25,

$\tilde{\xi}_{\alpha}=0$  & 0.5\tabularnewline
\bottomrule
\end{tabular}
\end{table}

Equations (\ref{eq:bt}-\ref{eq:St}) are solved numerically in the
non-dimensional form. As in the model of \citet{moss-sok08}, we assume
that the rotational shear is constant in latitude. The effect of differential
rotation is controlled by non-dimensional parameter $R_{\omega}={\displaystyle \frac{R^{2}}{\eta_{T}}\frac{\partial\Omega}{\partial r}}$,
the $\alpha$-effect is measured by parameter $R_{\alpha}={\displaystyle \frac{R\overline{\alpha}}{\eta_{T}}}$,
the magnetic buoyancy depends on $R_{\beta}={\displaystyle \frac{R}{\eta_{T}}\frac{\alpha_{MLT}u'}{\gamma}}$,
and the magnetic field is measured relative to the equipartition strength
$\mathrm{B_{eq}}=\sqrt{4\pi\overline{\rho}u'^{2}}$. Following \citet{ps11}
we put $R_{\omega}={\displaystyle \frac{R\Omega}{\eta_{T}}=}10^{3}$,
$R_{\alpha}=1$. This choice describes the $\alpha^{2}\Omega$ dynamo
regime with differential rotation as the main driver of axisymmetric
toroidal magnetic field. The non-axisymmetric modes do not take part
in the dynamo unless some non-axisymmetric phenomena come into the
play. To estimate the magnetic buoyancy parameter we employ results
of \citet{kp93} who argued that the maximum buoyancy velocity of
large-scale magnetic field of equipartition strength $B_{eq}$ is
of the order of 6 m/s. In the solar conditions, the magnetic diffusion
$\eta_{T}=10^{12}$cm$^{2}/$s \citep{1993AA...274..521M,2011SoPh..269....3R},
and $R_{\beta}\approx500$. In our models, the large-scale magnetic
field strength is below $B_{eq}$. Hence, we use by an order of magnitude
smaller value: $R_{\beta}=50$. In addition to nonlinear quenching
of the $\alpha$-effect, the magnetic buoyancy also causes a nonlinear
saturation of the dynamo process.

Table \ref{tab:C} shows the variable parameters of our models: $C_{\beta}$
controls active region emergence by means of the Parker's magnetic
buoyancy instability; $\sigma_{\xi}$ is the STD of the random $\alpha$
effect; $\tau_{\xi}$ is the correlation time of the random $\alpha$
fluctuations, measured relative to dynamo period $P$. In models M2d
and M3b, non-axisymmetric perturbations of $\alpha$ are neglected:
$\tilde{\xi}_{\alpha}=0$ (however, the axisymmetric $\alpha$ component
fluctuates). For the chosen set of parameters, the antisymmetric parity
is dominant in stationary dynamo regimes of models M0 and M1a. The
magnetic field distribution of model M1a in the stationary stage is
used as the initial condition for subsequent runs.

\begin{figure}
\centering \includegraphics[width=0.8\columnwidth]{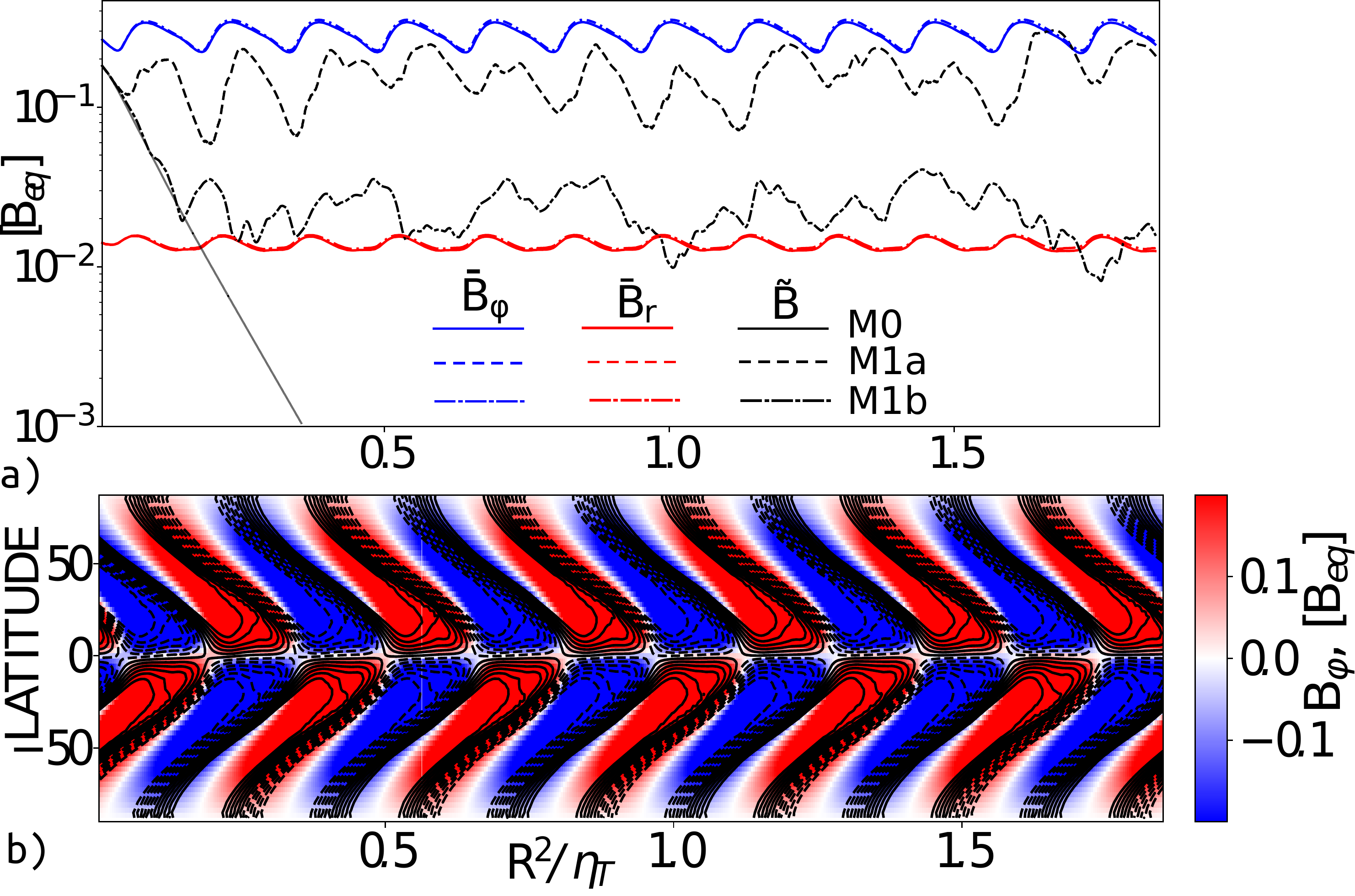} \caption{\label{fig:M0}{\bf a) 
Comparison of models M0 (solid lines), M1a (dash lines) and M1b (dash-dot lines). $\overline{B}_{\phi}$ (blue) and $\overline{B}_{r}$(red) are the mean axisymmetric toroidal
and radial magnetic field, $\tilde{B}$(black) is the mean
non-axisymmetric magnetic field in the units of the equipartition field strength. The parameters of the axisymmetric magnetic field in the models are very similar.} b) The time-latitude diagram of the toroidal
magnetic field (background image) and the radial magnetic field (contours
are in the range of $\pm 0.002$) for model M1a.}
\end{figure}

\section{Numerical Results}

Figure \ref{fig:M0} shows evolution of dynamo properties in the reference
models, M0 and M1a,b. The mean toroidal magnetic field varies around
$0.4B_{eq}$, while the radial magnetic field is by a factor of 50
smaller. Model M0 shows that the non-axisymmetric magnetic field is
decaying with time. Model M1a maintains large-scale non-axisymmetric
magnetic field which is seeded by diffusive dispersion of bipolar
active regions. The mean strength of the non-axisymmetric magnetic
field is a bit less or around the mean strength of the axisymmetric
toroidal magnetic field. Interesting that the non-axisymmetric perturbations
almost do not affect the axisymmetric magnetic field evolution. However,
randomness of the active region formation results in some randomness
of the overall magnetic activity. 

Time-latitude diagrams of the axisymmetric toroidal and radial magnetic
fields for models M0 and M1a,b are very similar. Figure \ref{fig:M0}b
shows the diagrams for model M1a. We see that the model correctly
reproduces the solar-like dynamo waves of toroidal magnetic field
which drifts to the equator during the magnetic cycle. However, the
time-latitude evolution of radial magnetic field does not show the
polar branch (cf., \citealt{sten88}). Also, the phase relation between
the radial and toroidal components does not correspond to observations.
The observations, e.g., the above cited paper and \citet{2013AARv2166S}
or \citet{bl-br2003}, show that $\overline{B}_{r}\overline{B}_{\phi}<0$
at low latitudes of the Sun.

\begin{figure}
\centering \includegraphics[width=0.95\columnwidth]{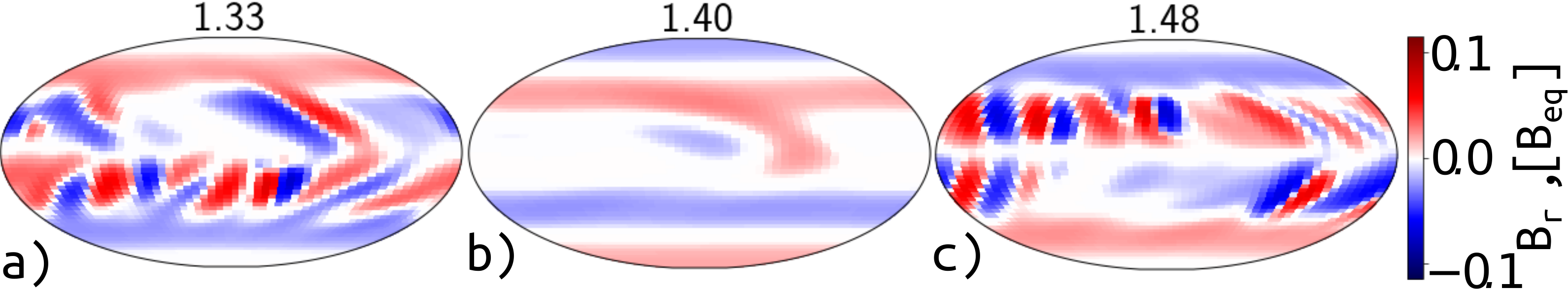} \caption{\label{fig:m1asn}Snapshots of the radial magnetic field distribution
in model M1a.}
\end{figure}

Figure \ref{fig:m1asn} shows snapshots of the radial magnetic field
distribution for different phases of the magnetic cycle in model M1a.
These snapshots show some solar-like features in organization of the
bipolar regions, such as the hemispheric Hale polarity rule. It is
driven by the axisymmetric toroidal magnetic field. Diffusive decay
of the bipolar regions and the differential rotation produce large-scale
unipolar regions which may extend from one hemisphere to another,
as seen in Figure \ref{fig:m1asn}a showing a snapshot for the magnetic
cycle maximum. Figure \ref{fig:m1asn}b shows that some remnants of
the bipolar regions can survive during the magnetic cycle minimum. 

Model M1b shows (Figure \ref{fig:M0}a, dash-dot line) that short-term
stochastic fluctuations of the $\alpha$ effect with $\sigma_{\xi}=0.25$
generate an order of magnitude smaller non-axisymmetric magnetic field
than the bipolar regions in model M1a. Model M1a shows the magnetic
cycle modulation of the mean strength of the non-axisymmetric magnetic
field. This feature is seen in the observations (see Sec. \ref{sec:observations}).
But, it is absent in the stochastic non-axisymmetric dynamo models.
{The non-axisymmetric dynamo is sustained by either non-axisymmetric magnetic buoyancy instability (e.g., model M1a) or the non-axisymmetric perturbations of the $\alpha$ effect (model M1b). Model M0 shows that the nonaxisisymmetric magnetic field vanishes if these two effects are simultaneously neglected. The large-scale non-axisymmetric magnetic field in model M1b is sustained by the non-axisymmetric fluctuations of the $\alpha$ effect (cf the results of model M0). In this case, the magnetic buoyancy of the axisymmetric magnetic field does not significantly affect the results. Also, the nonlinear interaction between the non-axisymmetric modes due to magnetic buoyancy is small because the strength of the non-axisymmetric components is well below the equipartition value.}

\begin{figure}
\centering \includegraphics[width=0.95\columnwidth]{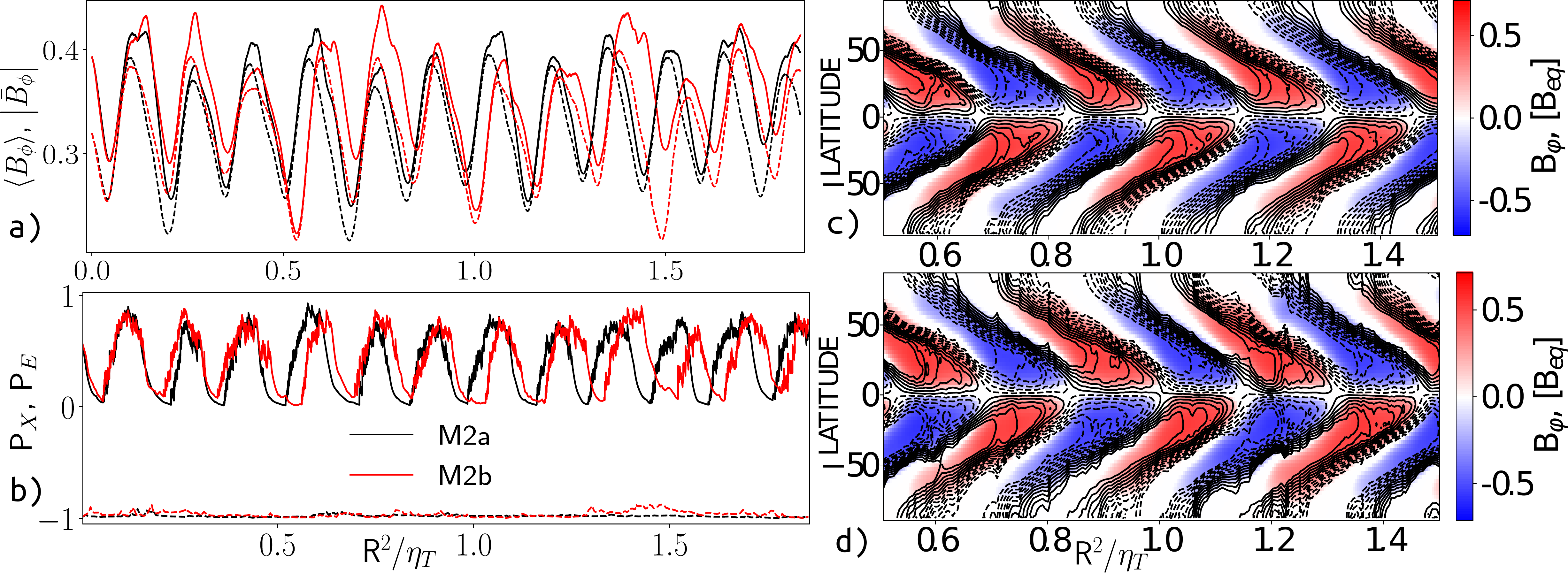} \caption{\label{fig:m2ab} Results for models M2a (black lines) and M2b (red
lines): a) the mean strength of the total toroidal magnetic field
(solid lines) and the axisymmetric toroidal magnetic field (dashed
lines); b) parameters $P_{X}$ (solid lines) and $P_{E}$ (dashed
lines). The time-latitude diagram of the axisymmetric toroidal magnetic
field (background image) and the radial magnetic field (contours are
plotted in the range of $\pm0.002$) for: c) model M2a; d) model M2b.}
\end{figure}

\begin{figure}
\centering \includegraphics[width=0.5\columnwidth]{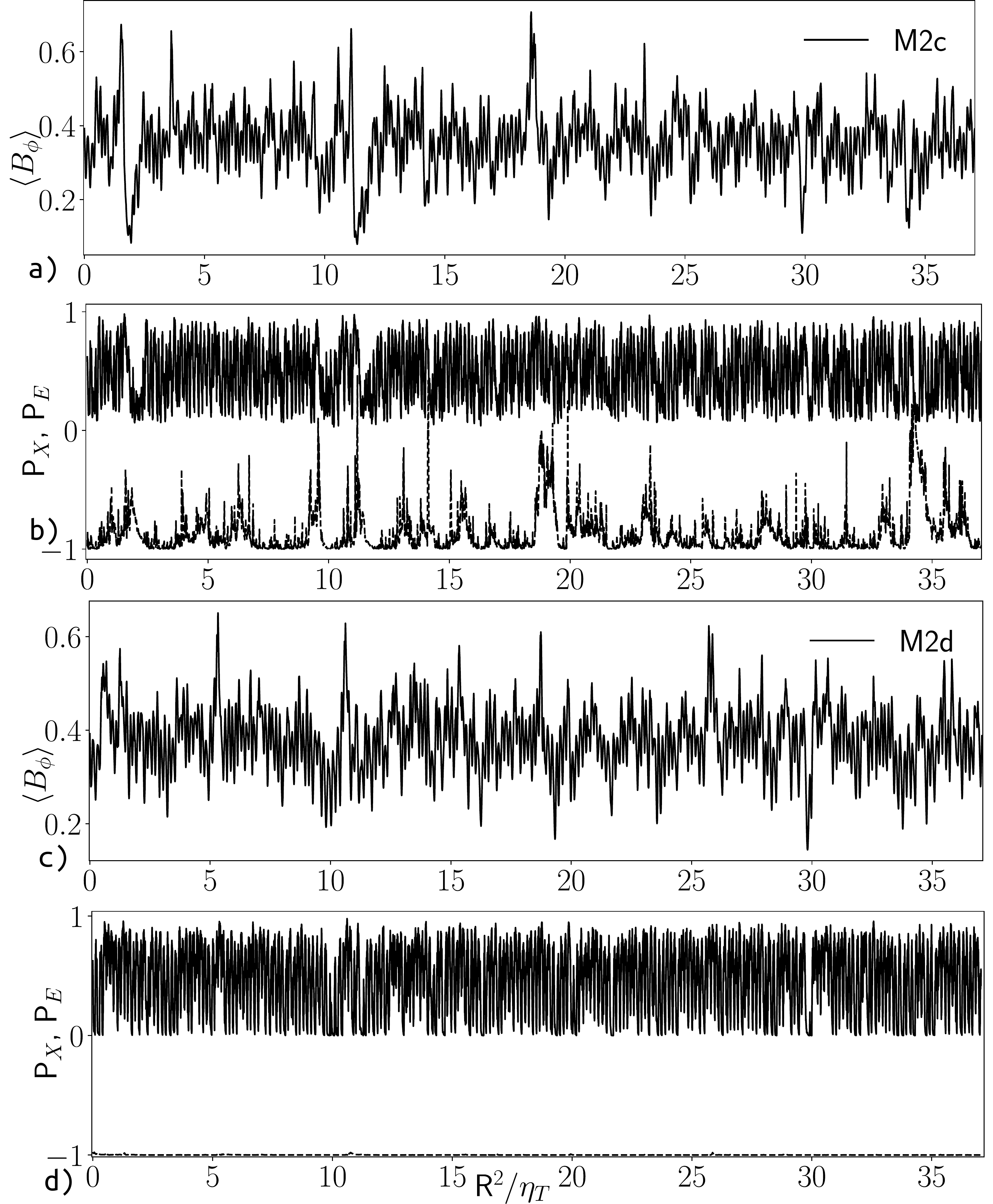} \caption{\label{fig:m2cd} The mean strength of the total toroidal magnetic
field (solid lines) for: a) model M2c and c) model M2d. Parameters
$P_{X}$ (solid lines) and $P_{E}$ (dashed lines) for: b) model M2c
and d) model M2d.}
\end{figure}

Figure \ref{fig:m2ab} shows results for the dynamo models which combine
effects of the bipolar regions and stochastic non-axisymmetric dynamo.
Models M2a and M2b employ a moderate level of short-term $\alpha$
fluctuations with $\sigma_{\xi}=0.25$ and $\sigma_{\xi}=0.5$ respectively.
In both cases, perturbations of the axisymmetric dynamo mode are not
strong. This is compatible with the results of the standard 1D models
(e.g., \citealp{choud92,h93,moss-sok08}), which showed that small
short-term fluctuations of the $\alpha$-effect in $\alpha\Omega$
dynamo models do not lead to substantial variations of the magnetic
cycle. Contrary, Figures~\ref{fig:m2ab}c and d show that the radial
magnetic field experiences strong noisy perturbations. In particular,
model M2b shows this in both the equatorial and radial regions. The
degree of non-axisymmetry of magnetic activity, parameter $P_{X}$,
varies from about zero during the minima to $0.95$ during the maxima
of magnetic cycles. It is found that with the increase of $\sigma_{\xi}$
the amount of non-axisymmetric magnetic field during the cycle minimum
increases. Another interesting effect is that the combined action
of the bipolar regions and non-axisymmetric perturbations of the $\alpha$-effect
can result in magnetic parity violations. Model M2b shows stronger
deviation of parameter $P_{E}$ from $-1$ than model M2a. The effect
becomes clear in model M2c with strong fluctuations of the $\alpha$-effect,
$\sigma_{\xi}=1$. 
\begin{figure}
\centering \includegraphics[width=0.5\columnwidth]{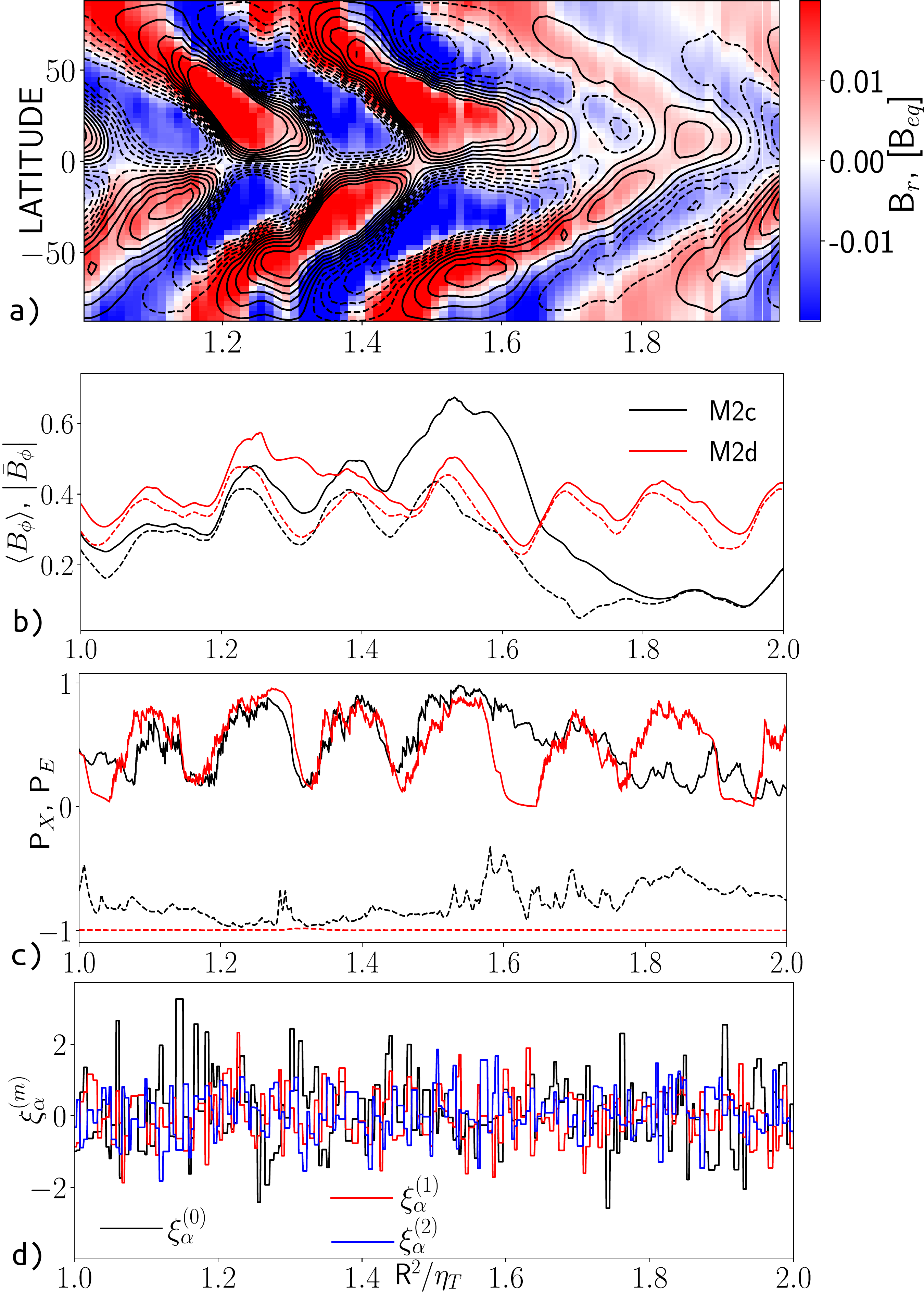} \caption{\label{fig:SC}a) The time-latitude diagram of the axisymmetric toroidal
magnetic field (background image) and the radial magnetic field (contour
are plotted in the range of $\pm0.002$) for model M2c during the
super-cycle event. b) The mean strength of the total toroidal magnetic
field (solid lines) for models M2c (black) and model M2d (red) for
a time interval which include the super-cycle events. c) Parameters
$P_{X}$ (solid lines) and $P_{E}$ (dashed lines) for these models;
{d) random fluctuations of the axisymmetric part of the $\alpha$
effect (black line) and the real part of the non-axisymmetric fluctuations
of the modes $m=1$ and $m=2$.}}
\end{figure}

This effect is demonstrated in Figure \ref{fig:m2cd}. Model M2c shows
strong cyclic variations of the mean total strength of toroidal magnetic
field. In some cases the two subsequent cycles are merged in one super
cycle. In such cycle, the total magnetic field strength exceeds the
strength of the background toroidal magnetic field by a factor of
2. The basal level of non-axisymmetric magnetic field is well above
zero especially in the activity minima before and after the super
cycle. Perturbations of the magnetic parity in model M2c are much
higher than in models M2a and M2b. Model M2d is a run without non-axisymmetric
fluctuations of the $\alpha$ effect, but with the same random realization
of axisymmetric fluctuations of the $\alpha$ effect as in model M2c.
Figure \ref{fig:m2cd}a shows evolution of $\overline{\left\langle B_{\phi}\right\rangle }$
in models M2c and M2d deviate from each other shortly after beginning.
Variations of parameters $P_{X}$ and $P_{E}$ differ substantially
in models M2c and M2d. In particular, $P_{E}\approx-1$ in model M2d,
but in model M2c parameter $P_{E}$ substantially deviates from $-1$
indicating asymmetry of the hemispheric magnetic activity.

Another interesting feature of model M2c is the super-cycle event
at the time moment, $R^{2}/\eta_{T}\approx1.5$ (Fig.~\ref{fig:m2cd}a).
It is illustrated in Figure \ref{fig:SC}. It is found that the super-cycle
is formed by merging a strong cycle with a weak cycle that occurred
during the decaying phase of the strong cycle. The super-cycle is
characterized by a high level of non-axisymmetric magnetic fields
activity. Also, the hemispheric asymmetry of magnetic field distributions
is high during the decaying phase of the cycle.

\begin{figure}
\centering \includegraphics[width=0.95\columnwidth]{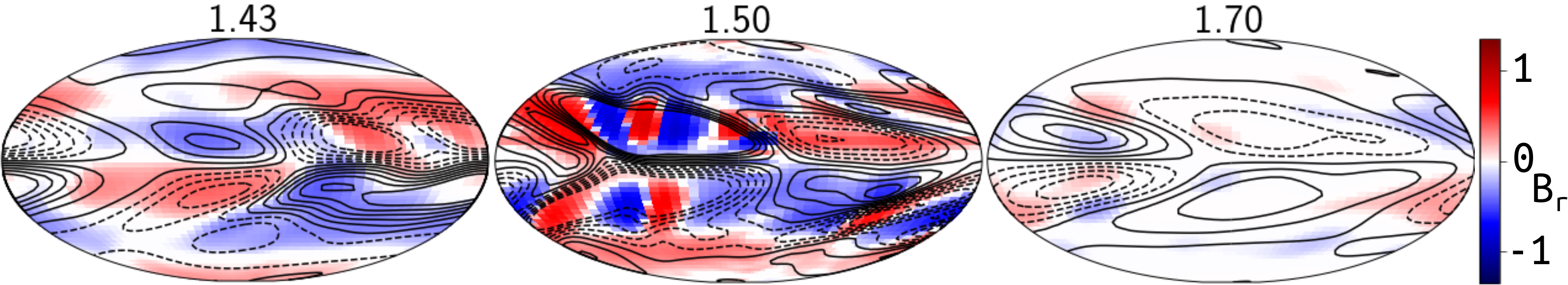} \caption{\label{fig:m2csn}Snapshots of the radial magnetic field distribution
in model M2c, contours show the $B_{\phi}$ component in range of
$\pm B_{eq}$.}
\end{figure}

Figures \ref{fig:m2csn} and \ref{fig:SC}b,c show that the super
cycle starts from quite strong ($\left\langle B_{r}\right\rangle \approx0.5B_{eq}$)
and also non-axisymmetric distributions of magnetic field, which is
generated during the preceding minimum of magnetic activity. During
the cycle maximum, superposition of the strong axisymmetric magnetic
field and magnetic field of new bipolar regions produces wide magnetic
`nests' with the field strength: $\left\langle B_{r}\right\rangle >B_{eq}$.
At the end of the super-cycle, the magnetic field remains non-axisymmetric,
and is also antisymmetric relative to the equator. Simultaneously,
the strength of the axisymmetric toroidal magnetic field reaches the
deep minimum, see Figure \ref{fig:SC}b. Remarkably, model M2d does
not show similar events. 
{Figure \ref{fig:SC}d shows fluctuations
of the axisymmetric and non-axisymmetric parts of the $\alpha$-effect.
The period preceding the super-cycle event is characterized by a
moderate level of the random fluctuations. The mean $\alpha$-effect
changed its sign from positive to negative for a short period around time $R^{2}/\eta_{T}\approx1.25$.
It is unclear if this event triggered the dramatic increase of activity
in the subsequent two magnetic cycles. }

\begin{figure}
\centering \includegraphics[width=0.75\columnwidth]{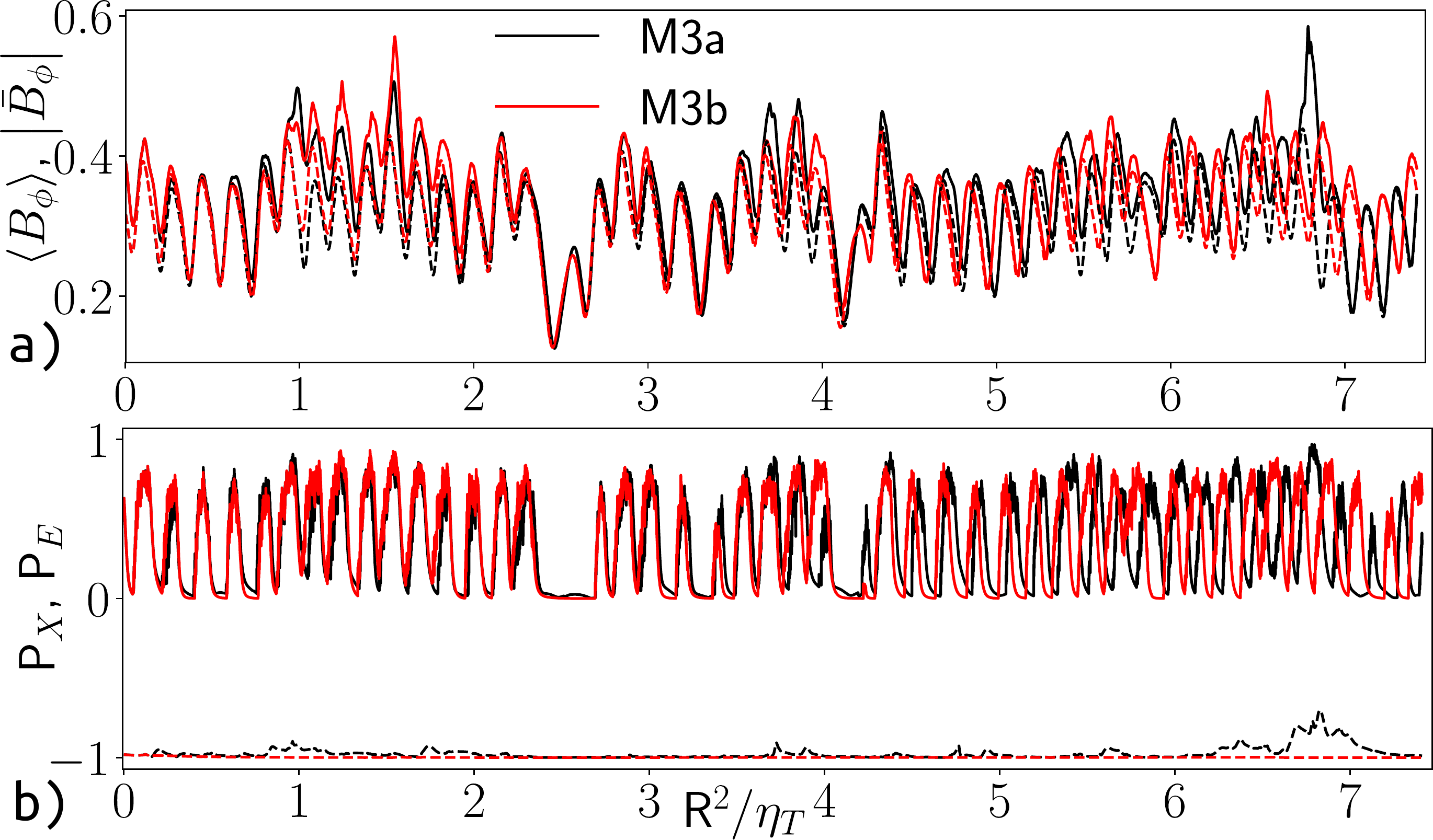} \caption{\label{fig:m3ab}The same as Figure \ref{fig:m2ab} a and b for models
M3a and M3b}
\end{figure}

The case of small long-term fluctuations of $\alpha$ is considered
in models M3a and M3b, see Figure~\ref{fig:m3ab}. These models have
very similar time evolutions. The time interval in the model runs
covers the period of about fifty cycles, which includes two deep minima
of magnetic activity. In model M3a, there is one event which is similar
to the super cycle of model M2c. Both models M3a and M3b show a high
level of non-axisymmetry during the cycle maxima. In model M3b the
non-axisymmetric dynamo is suppressed. This run shows strictly antisymmetric
hemispheric magnetic activity. Variations of parity index $P_{E}$
in model M3a are similar to those in model M2c. The higher magnitude
of the non-axisymmetric variations of $\alpha$, the higher parity
index $P_{E}$.

\section{Comparison with Observational Data}

\label{sec:observations} 
\begin{figure}
\centering \includegraphics[width=0.85\columnwidth]{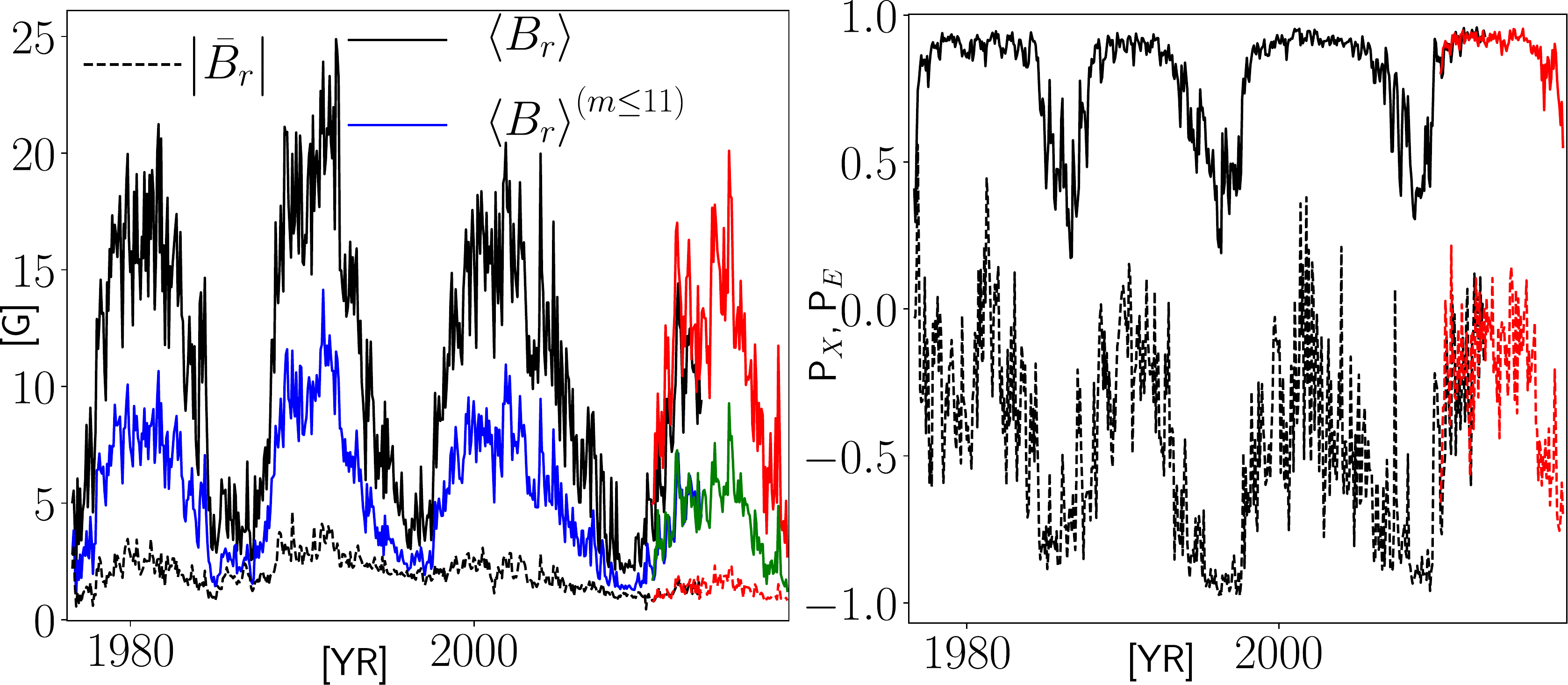} \caption{\label{fig:mdi}a) The unsigned radial magnetic field of the Sun from
synoptic magnetograms; b) The non-axisymmetry index, $P_{X}$ (solid
line); The equatorial symmetry index of axisymmetric magnetic field,
$P_{E}$(dashed line).}
\end{figure}

\begin{figure}
\centering \includegraphics[width=0.75\columnwidth]{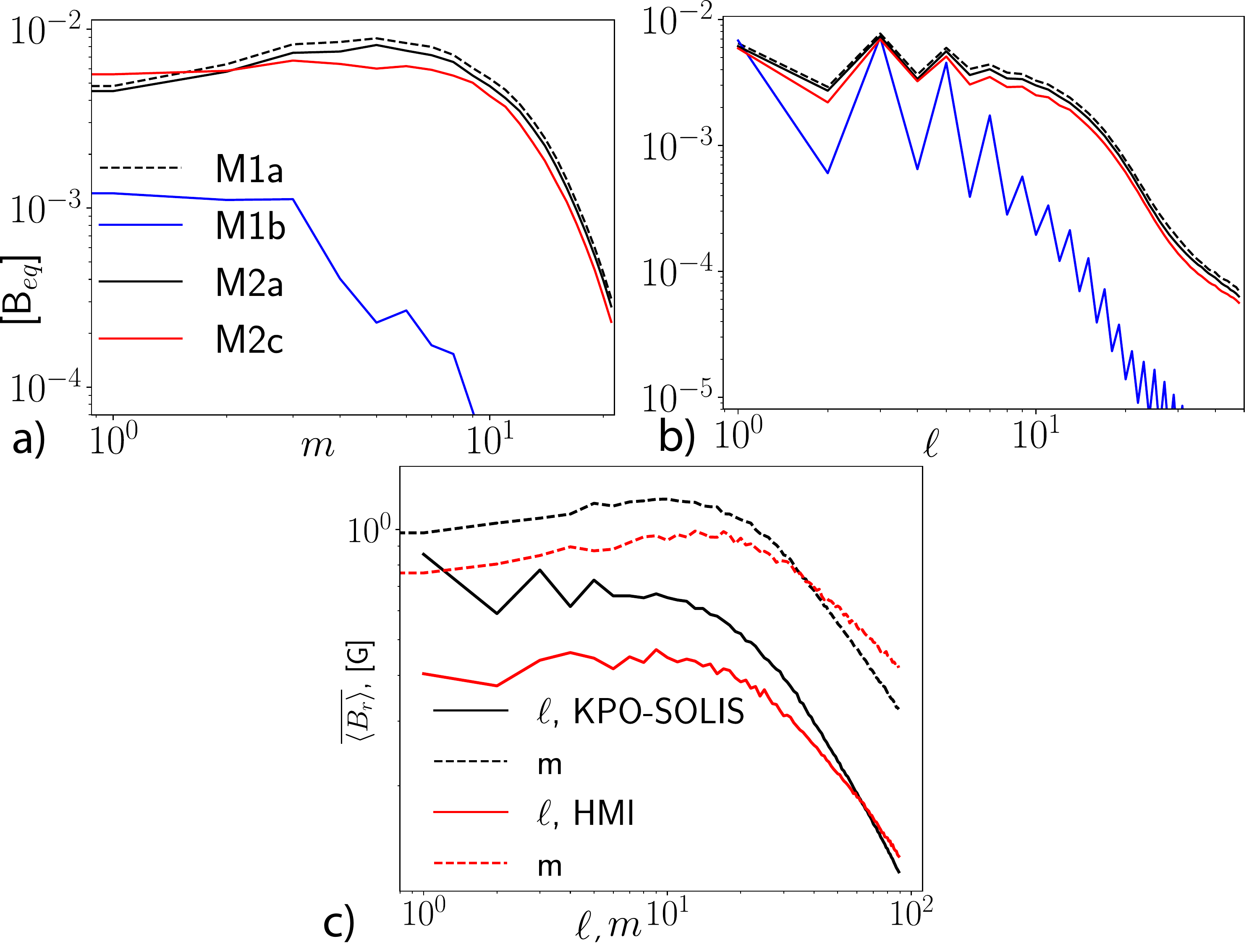} \caption{\label{fig:spec}a) The mean (over latitudes and time) spectral density
flux of radial magnetic field as a function of azimuthal order $m$;
b) the mean $\ell$-spectra of the radial magnetic field for the dynamo
models.}
\end{figure}

To compare our results with observations we use synoptic maps of radial
magnetic field from the KPO, SOLIS and SDO/HMI data archives \citep{Harvey1980,Bertello2014,Scherrer2012}.
Using the synoptic maps we calculate the surface mean of the unsigned
radial magnetic field, $\overline{\left|\left\langle \mathbf{B}_{r}\right\rangle \right|}$
and the same for the axisymmetric radial magnetic field, $\overline{\left|\overline{\mathbf{B}}_{r}\right|}$,
the parity index $P_{E}$, the level of non-axisymmetry $P_{X}$,
see, Eqs (\ref{eq:par}) and (\ref{eq:nx}). To calculate parameter
$P_{X}$ we perform the spherical harmonic decomposition up to $m=10$.
Results are shown in Figure \ref{fig:mdi}. In agreement with the
results of \citet{2013AARv2166S} $\overline{\left|\left\langle \mathbf{B}_{r}\right\rangle \right|}$
reaches about 20-25 G at the solar maxima. 
The strength of large-scale magnetic field is about a factor of two
smaller than the total magnetic field strength. Figure~\ref{fig:mdi}b
shows that $P_{X}\approx1$ during the solar maxima, and $P_{X}\approx0.5$
during the cycle minima. This means that the total basal level of
the $m=1\dots11$ modes of large-scale non-axisymmetric magnetic field
exceeds the basal level of the axisymmetric component of magnetic
field. Parity of the axisymmetric radial magnetic field varies from
$P_{E}\approx0$ during the active phase of the solar cycle to $P_{E}\approx-1$
during the cycle minimum.

Figure \ref{fig:spec} shows mean spectra of the azimuthal magnetic
field distributions in our models and in the observations. Model M1a
(non-axisymmetric dynamo is suppressed) shows maximum for the modes
with $m$=5 and 6. It shows a slow decrease towards small $m$ and
fast decrease for high $m$. The spectrum is formed by diffusive decay
of the bipolar magnetic active regions. The low-$m$ branch gradually
disappears with increase of the non-axisymmetric $\alpha$-effect
perturbations. Model M1b shows a monotonic decrease of the mean magnetic
field strength with the increase of $m$. To compare with observations
we apply the FFT transform to the set of synoptic magnetograms of
the radial magnetic fields. Additionally, we scale the results of
our runs for $\left\langle B_{r}\right\rangle $ in units of $B_{eq}$
by a factor of 200. Figure \ref{fig:spec} c shows comparison of the
models with observations. The observational data show the maximum
of the mean magnetic field strength around $m=10$. The decrease towards
low $m$ is not as strong as in model M1a, but it is similar to models
M2b and and M2c. We interpret this as the presence of stochastic non-axisymmetric
dynamo action. In the observational data, diffusive decay of the high
$m$-modes is less prominent than in our models. This can be interpreted
in two different ways. First, the diffusion coefficient in the model
may be too high. Second, the solar dynamo can be multiscale and intermittent
phenomenon. These properties are not captured in our models.

Figures \ref{fig:spec}b and d show the normalized (by factor $2\ell+1$)
$\ell$-spectra of spherical harmonic decomposition of radial magnetic
field for the models and observations. The observational data are
processed using the Python interface of SHTools library (https://shtools.oca.eu/shtools/index.html).
Both models and observations show a maximum for the modes of $\ell=3,5$
and dips for low-degree even modes. This is in agreement with analysis
of \citet{sten88,2013AARv2166S}. As for the $m$-spectra, the magnetic
field distribution in the models occupies a smaller spectral interval
than in the real data.

\section{Discussion}

We used a relatively simple 2D mean-field dynamo model to understand
whether non-axisymmetric dynamo modes can be generated and maintained
in the presence of random non-axisymmetric perturbations of the $\alpha$-effect
and due to diffusive decay of emerging bipolar regions. Without such
perturbations the non-axisymmetric dynamo modes do not develop, and
the dynamo solutions are axisymmetric. Here, for the first
time, we demonstrate that the non-axisymmetric dynamo can happen even
for the solar conditions, {\bf despite the strong differential rotation.} 

The reduced 2D non-linear non-axisymmetric dynamo models considered
in this paper allow us to investigate influence of the perturbations
on properties of the dynamo cycles, including the degree of non-axisymmetry
of the dynamo-generated magnetic field and the hemispheric asymmetry
in different phases of the magnetic cycles. Despite the simplicity,
the models give insight on how large-scale non-axisymmetric magnetic
structures observed on the Sun may develop, as well as directions
for development of more realistic non-axisymmetric dynamo models.
Our model of emerging bipolar regions was based on parameterization
of the Parker's buoyancy instability of the axisymmetric toroidal
magnetic field. Short- and long-term stochastic large-scale (up to
octupole) perturbations of the $\alpha$-effect of various amplitude
were considered separately and in combination with the bipolar regions.

It is found that large-scale non-axisymmetric dynamo modes can be
excited and maintain because of diffusion of emerging bipolar magnetic
regions. Without perturbations of the $\alpha$-effect the non-axisymmetric
magnetic field evolution affects evolution of the axisymmetric toroidal
magnetic field via magnetic buoyancy. This kind of coupling was discussed
previously by \citet{pk15}. In general, formation of large-scale
non-axisymmetric magnetic field due to active region decay is usually
accepted for granted \citep{2012LRSP96M}. Yet, the origin of the
large-scale non-axisymmetric component during the solar minima is
poorly understood. Our results show that remnants of decaying bipolar
regions can persists during the solar minima when the cycle duration
is determined by the turbulent diffusion scale.

It is found that short-term stochastic non-axisymmetric fluctuations
of the $\alpha$-effect with the standard deviation of 25\% relative
to the axisymmetric level ($\sigma_{\xi}=0.25$) ) can generate weak
non-axisymmetric magnetic field. Its strength is two orders of magnitude
smaller than the strength of magnetic field in magnetic bipolar regions.
The magnitude of turbulent $\alpha$-effect fluctuations is unknown.
\citet{h93,oss-h96a} suggested that a high level short-term fluctuations
with $\sigma_{\xi}=1$ can explain the Grand minima events. The results
of \citet{moss-sok08} showed that a low level ($\sigma_{\xi}=0.1$)
of long-term $\alpha$ fluctuations is another option. Our model with
strong non-axisymmetric fluctuations of the $\alpha$-effect ($\sigma_{\xi}=1$)
shows super-cycle events caused by a non-linear interaction of the
non-axisymmetric dynamo and the process of formation of bipolar regions.
The super-cycle magnitude is more than two times greater than the
mean maximum of the magnetic cycles. The model run lasted 250 cycles
(Fig.~\ref{fig:m2cd}a,b) shows a few other prolonged cycles of somewhat
smaller magnitudes, as well as periods of low magnetic activity, but
the super-cycle events are rare.

Comparison the mean spectra of the latitudinally averaged strength
of large-scale radial magnetic field with synoptic observations of
solar magnetic fields during the last four cycles showed that the
short-term non-axisymmetric perturbations of the $\alpha$-effect
with $\sigma_{\xi}=0.5-1$ can be an option to explain the low azimuthal
order part of the spectrum. Our simulations show that without the
non-axisymmetric dynamo the magnetic field strength in this part of
spectrum would be factor 2-3 lower than it is seen in the observational
data. In comparison of our results with solar observations, we have
to keep in mind that in our model the toroidal magnetic field generation
and the bipolar region formation occur in the same place. {Taking
into account the strong effect of the differential rotation in our
models and the dominant role of the axisymmetric toroidal magnetic
field, the shallow surface tachocline could be considered as a relevant
place for our models. This can be different from the solar case \citep{b05}.
Also, our results can be applied to stars with shallow convection
zones (late-F and early-G spectral classes).}

Young solar analogs often show a combination of the `inactive' and
`active' branches of the cyclic activity \citet{2009AA_strassm}.
The active branch shows long cycles \citep{1999ApJ_sa_br,2007ApJ_bohm}.
\citet{2016ApJ823.133P} conjectured that the active branch can be
due to the non-axisymmetric dynamo. This conclusion is supported by
results of \citet{2016MNRAS1129S}. Here, for the first time we demonstrate
the non-axisymmetric dynamo for a solar-type model.

\acknowledgements{VVP conducted this work as a part of FR II.16 of ISTP SB RAS. AGK
thanks for support the NASA Grants NNX14AB68G, and NNX16AP05H. The
dynamo code as well as codes to process the observational data are
provided online by \citet{2dspy}}

\bibliographystyle{aasjournal}


\begin{thebibliography}{}
\expandafter\ifx\csname natexlab\endcsname\relax\def\natexlab#1{#1}\fi
\providecommand{\url}[1]{\href{#1}{#1}}
\providecommand{\dodoi}[1]{doi:~\href{http://doi.org/#1}{\nolinkurl{#1}}}
\providecommand{\doeprint}[1]{\href{http://ascl.net/#1}{\nolinkurl{http://ascl.net/#1}}}
\providecommand{\doarXiv}[1]{\href{https://arxiv.org/abs/#1}{\nolinkurl{https://arxiv.org/abs/#1}}}

\bibitem[{{Berdyugina} {et~al.}(2006){Berdyugina}, {Moss}, {Sokoloff}, \&
  {Usoskin}}]{berd06}
{Berdyugina}, S.~V., {Moss}, D., {Sokoloff}, D., \& {Usoskin}, I.~G. 2006,
  \aap, 445, 703, \dodoi{10.1051/0004-6361:20053454}

\bibitem[{{Bertello} {et~al.}(2014){Bertello}, {Pevtsov}, {Petrie}, \&
  {Keys}}]{Bertello2014}
{Bertello}, L., {Pevtsov}, A.~A., {Petrie}, G.~J.~D., \& {Keys}, D. 2014,
  \solphys, 289, 2419, \dodoi{10.1007/s11207-014-0480-3}

\bibitem[{{Bigazzi} \& {Ruzmaikin}(2004)}]{bigruz}
{Bigazzi}, A., \& {Ruzmaikin}, A. 2004, \apj, 604, 944, \dodoi{10.1086/381932}

\bibitem[{{Blackman} \& {Brandenburg}(2003)}]{bl-br2003}
{Blackman}, E.~G., \& {Brandenburg}, A. 2003, \apjl, 584, L99,
  \dodoi{10.1086/368374}

\bibitem[{{B{\"o}hm-Vitense}(2007)}]{2007ApJ_bohm}
{B{\"o}hm-Vitense}, E. 2007, \apj, 657, 486, \dodoi{10.1086/510482}

\bibitem[{Brandenburg(2005)}]{b05}
Brandenburg, A. 2005, Astrophys. J., 625, 539

\bibitem[{{Brandenburg} {et~al.}(2013){Brandenburg}, {Kleeorin}, \&
  {Rogachevskii}}]{2013ApJ776L23B}
{Brandenburg}, A., {Kleeorin}, N., \& {Rogachevskii}, I. 2013, \apj, 776, L23,
  \dodoi{10.1088/2041-8205/776/2/L23}

\bibitem[{{Brandenburg} {et~al.}(1989){Brandenburg}, {Krause}, {Meinel},
  {Moss}, \& {Tuominen}}]{bran89}
{Brandenburg}, A., {Krause}, F., {Meinel}, R., {Moss}, D., \& {Tuominen}, I.
  1989, \aap, 213, 411

\bibitem[{{Brandenburg} \& {Subramanian}(2005)}]{brsu05}
{Brandenburg}, A., \& {Subramanian}, K. 2005, \physrep, 417, 1,
  \dodoi{10.1016/j.physrep.2005.06.005}

\bibitem[{{Cameron} \& {Sch{\"u}ssler}(2017)}]{2017ApJ843.111C}
{Cameron}, R.~H., \& {Sch{\"u}ssler}, M. 2017, \apj, 843, 111,
  \dodoi{10.3847/1538-4357/aa767a}

\bibitem[{{Choudhuri}(1992)}]{choud92}
{Choudhuri}, A.~R. 1992, \aap, 253, 277

\bibitem[{{Getling}(2001)}]{2001ARep...45..569G}
{Getling}, A.~V. 2001, Astronomy Reports, 45, 569, \dodoi{10.1134/1.1383816}

\bibitem[{{Glencross}(1974)}]{1974Natur250.717G}
{Glencross}, W.~M. 1974, \nat, 250, 717, \dodoi{10.1038/250717a0}

\bibitem[{{Harvey} {et~al.}(1980){Harvey}, {Gillespie}, {Miedaner}, \&
  {Slaughter}}]{Harvey1980}
{Harvey}, J., {Gillespie}, B., {Miedaner}, P., \& {Slaughter}, C. 1980, NASA
  STI/Recon Technical Report N, 81

\bibitem[{{Hoyng}(1993)}]{h93}
{Hoyng}, P. 1993, \aap, 272, 321

\bibitem[{{Jennings} {et~al.}(1990){Jennings}, {Brandenburg}, {Tuominen}, \&
  {Moss}}]{1990AA230.463J}
{Jennings}, R., {Brandenburg}, A., {Tuominen}, I., \& {Moss}, D. 1990, \aap,
  230, 463

\bibitem[{{Kitchatinov} \& {Pipin}(1993)}]{kp93}
{Kitchatinov}, L.~L., \& {Pipin}, V.~V. 1993, \aap, 274, 647


\bibitem[{{Kitiashvili} {et~al.}(2010){Kitiashvili}, {Kosovichev}, {Wray}, \&
  {Mansour}}]{2010ApJ719.307K}
{Kitiashvili}, I.~N., {Kosovichev}, A.~G., {Wray}, A.~A., \& {Mansour}, N.~N.
  2010, \apj, 719, 307, \dodoi{10.1088/0004-637X/719/1/307}

\bibitem[{Krause \& R\"adler(1980)}]{KR80}
Krause, F., \& R\"adler, K.-H. 1980, Mean-Field Magnetohydrodynamics and Dynamo
  Theory (Berlin: Akademie-Verlag), 271

\bibitem[{{Kuzanyan}(1998)}]{ku98}
{Kuzanyan}, K.~M. 1998, in Astronomical Society of the Pacific Conference
  Series, Vol. 154, Cool Stars, Stellar Systems, and the Sun, ed. R.~A.
  {Donahue} \& J.~A. {Bookbinder}, 1286

\bibitem[{{Leka} {et~al.}(2013){Leka}, {Barnes}, {Birch}, {Gonzalez-
  Hernandez}, {Dunn}, {Javornik}, \& {Braun}}]{2013ApJ762.130L}
{Leka}, K.~D., {Barnes}, G., {Birch}, A.~C., {et~al.} 2013, \apj, 762, 130,
  \dodoi{10.1088/0004-637X/762/2/130}

\bibitem[{{Losada} {et~al.}(2017){Losada}, {Warnecke}, {Glogowski}, {Roth},
  {Brandenburg}, {Kleeorin}, \& {Rogachevskii}}]{2017IAUS327.46L}
{Losada}, I.~R., {Warnecke}, J., {Glogowski}, K., {et~al.} 2017, in IAU
  Symposium, Vol. 327, Fine Structure and Dynamics of the Solar Atmosphere, ed.
  S.~{Vargas Dom{\'{\i}}nguez}, A.~G. {Kosovichev}, P.~{Antolin}, \&
  L.~{Harra}, 46--59

\bibitem[{{Mackay} \& {Yeates}(2012)}]{2012LRSP96M}
{Mackay}, D.~H., \& {Yeates}, A.~R. 2012, Living Reviews in Solar Physics, 9,
  6, \dodoi{10.12942/lrsp-2012-6}

\bibitem[{{Martin}(2018)}]{2018FrASS517M}
{Martin}, S.~F. 2018, Frontiers in Astronomy and Space Sciences, 5, 17,
  \dodoi{10.3389/fspas.2018.00017}

\bibitem[{{Martinez Pillet} {et~al.}(1993){Martinez Pillet}, {Moreno-Insertis},
  \& {Vazquez}}]{1993AA...274..521M}
{Martinez Pillet}, V., {Moreno-Insertis}, F., \& {Vazquez}, M. 1993, \aap, 274,
  521

\bibitem[{{Moss}(1999)}]{moss99}
{Moss}, D. 1999, \mnras, 306, 300, \dodoi{10.1046/j.1365-8711.1999.02510.x}

\bibitem[{{Moss} {et~al.}(2008){Moss}, {Sokoloff}, {Usoskin}, \&
  {Tutubalin}}]{moss-sok08}
{Moss}, D., {Sokoloff}, D., {Usoskin}, I., \& {Tutubalin}, V. 2008, Solar
  Phys., 250, 221

\bibitem[{{Moss} {et~al.}(1990){Moss}, {Tuominen}, \&
  {Brandenburg}}]{1990AA240.142M}
{Moss}, D., {Tuominen}, I., \& {Brandenburg}, A. 1990, \aap, 240, 142

\bibitem[{{Noyes} {et~al.}(1984){Noyes}, {Weiss}, \&
  {Vaughan}}]{1984ApJ...287..769N}
{Noyes}, R.~W., {Weiss}, N.~O., \& {Vaughan}, A.~H. 1984, \apj, 287, 769,
  \dodoi{10.1086/162735}

\bibitem[{{Ol{\'a}h} {et~al.}(2009){Ol{\'a}h}, {Koll{\'a}th}, {Granzer},
  {Strassmeier}, {Lanza}, {J{\"a}rvinen}, {Korhonen}, {Baliunas}, {Soon},
  {Messina}, \& {Cutispoto}}]{2009AA_strassm}
{Ol{\'a}h}, K., {Koll{\'a}th}, Z., {Granzer}, T., {et~al.} 2009, \aap, 501,
  703, \dodoi{10.1051/0004-6361/200811304}

\bibitem[{{Ossendrijver} \& {Hoyng}(1996)}]{oss-h96a}
{Ossendrijver}, A.~J.~H., \& {Hoyng}, P. 1996, \aap, 313, 959

\bibitem[{Parker(1979)}]{park}
Parker, E.~N. 1979, Cosmical magnetic fields: Their origin and their activity
  (Oxford: Clarendon Press)

\bibitem[{{Parker}(1984)}]{1984ApJ...281..839P}
{Parker}, E.~N. 1984, \apj, 281, 839, \dodoi{10.1086/162163}

\bibitem[{{Parker}(1993)}]{park93}
---. 1993, \apj, 408, 707, \dodoi{10.1086/172631}

\bibitem[{{Passos} {et~al.}(2014){Passos}, {Nandy}, {Hazra}, \&
  {Lopes}}]{2014AA563A18P}
{Passos}, D., {Nandy}, D., {Hazra}, S., \& {Lopes}, I. 2014, \aap, 563, A18,
  \dodoi{10.1051/0004-6361/201322635}

\bibitem[{Pipin(2018)}]{2dspy}
Pipin, V. 2018, VVpipin/2DSPDy 0.1.1, \dodoi{10.5281/zenodo.1413149}

\bibitem[{{Pipin}(2008)}]{pi08Gafd}
{Pipin}, V.~V. 2008, Geophysical and Astrophysical Fluid Dynamics, 102, 21

\bibitem[{{Pipin}(2017)}]{2017MNRAS.466.3007P}
---. 2017, \mnras, 466, 3007, \dodoi{10.1093/mnras/stw3182}

\bibitem[{Pipin \& Kosovichev(2015)}]{pk15}
Pipin, V.~V., \& Kosovichev, A.~G. 2015, The Astrophysical Journal, 813, 134

\bibitem[{{Pipin} \& {Kosovichev}(2016)}]{2016ApJ823.133P}
{Pipin}, V.~V., \& {Kosovichev}, A.~G. 2016, \apj, 823, 133,
  \dodoi{10.3847/0004-637X/823/2/133}

\bibitem[{{Pipin} {et~al.}(2014){Pipin}, {Moss}, {Sokoloff}, \&
  {Hoeksema}}]{pial14}
{Pipin}, V.~V., {Moss}, D., {Sokoloff}, D., \& {Hoeksema}, J.~T. 2014, \aap,
  567, A90, \dodoi{10.1051/0004-6361/201323319}

\bibitem[{Pipin \& Sokoloff(2011)}]{ps11}
Pipin, V.~V., \& Sokoloff, D.~D. 2011, Physica Scripta, 84, 065903

\bibitem[{{Pipin} {et~al.}(2012){Pipin}, {Sokoloff}, \&
  {Usoskin}}]{pipea2012AA}
{Pipin}, V.~V., {Sokoloff}, D.~D., \& {Usoskin}, I.~G. 2012, \aap, 542, A26,
  \dodoi{10.1051/0004-6361/201118733}

\bibitem[{{Raedler}(1986)}]{rad86AN}
{Raedler}, K.-H. 1986, Astronomische Nachrichten, 307, 89,
  \dodoi{10.1002/asna.2113070205}

\bibitem[{{Raedler} {et~al.}(1990){Raedler}, {Wiedemann}, {Brandenburg},
  {Meinel}, \& {Tuominen}}]{radler90}
{Raedler}, K.-H., {Wiedemann}, E., {Brandenburg}, A., {Meinel}, R., \&
  {Tuominen}, I. 1990, \aap, 239, 413

\bibitem[{{R{\"u}diger} {et~al.}(2011){R{\"u}diger}, {Kitchatinov}, \&
  {Brandenburg}}]{2011SoPh..269....3R}
{R{\"u}diger}, G., {Kitchatinov}, L.~L., \& {Brandenburg}, A. 2011, \solphys,
  269, 3, \dodoi{10.1007/s11207-010-9683-4}

\bibitem[{{Saar} \& {Brandenburg}(1999)}]{1999ApJ_sa_br}
{Saar}, S.~H., \& {Brandenburg}, A. 1999, \apj, 524, 295,
  \dodoi{10.1086/307794}

\bibitem[{{Scherrer} {et~al.}(2012){Scherrer}, {Schou}, {Bush}, {Kosovichev},
  {Bogart}, {Hoeksema}, {Liu}, {Duvall}, {Zhao}, {Title}, {Schrijver},
  {Tarbell}, \& {Tomczyk}}]{Scherrer2012}
{Scherrer}, P.~H., {Schou}, J., {Bush}, R.~I., {et~al.} 2012, \solphys, 275,
  207, \dodoi{10.1007/s11207-011-9834-2}

\bibitem[{{See} {et~al.}(2016){See}, {Jardine}, {Vidotto}, {Donati}, {Boro
  Saikia}, {Bouvier}, {Fares}, {Folsom}, {Gregory}, {Hussain}, {Jeffers},
  {Marsden}, {Morin}, {Moutou}, {do Nascimento}, {Petit}, \&
  {Waite}}]{2016MNRAS1129S}
{See}, V., {Jardine}, M., {Vidotto}, A.~A., {et~al.} 2016, \mnras, 462, 4442,
  \dodoi{10.1093/mnras/stw2010}

\bibitem[{{Stein} \& {Nordlund}(2012)}]{2012ApJ753L13S}
{Stein}, R.~F., \& {Nordlund}, {\AA}. 2012, \apjl, 753, L13,
  \dodoi{10.1088/2041-8205/753/1/L13}

\bibitem[{{Stenflo}(2012)}]{2012AA547A93S}
{Stenflo}, J.~O. 2012, \aap, 547, A93, \dodoi{10.1051/0004-6361/201219833}

\bibitem[{{Stenflo}(2013)}]{2013AARv2166S}
---. 2013, \aapr, 21, 66, \dodoi{10.1007/s00159-013-0066-3}

\bibitem[{{Stenflo} \& {Guedel}(1988)}]{sten88}
{Stenflo}, J.~O., \& {Guedel}, M. 1988, \aap, 191, 137

\bibitem[{{Stix}(1977)}]{stix77}
{Stix}, M. 1977, \aap, 59, 73

\bibitem[{{Usoskin} {et~al.}(2009){Usoskin}, {Sokoloff}, \& {Moss}}]{uetal09}
{Usoskin}, I.~G., {Sokoloff}, D., \& {Moss}, D. 2009, \solphys, 254, 345,
  \dodoi{10.1007/s11207-008-9293-6}

\bibitem[{{Wang} \& {Sheeley}(1990)}]{1990ApJ355.726W}
{Wang}, Y.-M., \& {Sheeley}, Jr., N.~R. 1990, \apj, 355, 726,
  \dodoi{10.1086/168805}

\end{thebibliography}

\end{document}